# Ultra-confined $q^{2/3}$- plasmons and optical properties of graphene-transition metal dichalcogenide heterostructures

Partha Goswami

*D.B.College, University of Delhi, Kalkaji, New Delhi, India*

**Abstract** We report the interesting possibilities related to the plasmonics and the key parameters of optoelectronics, such as the absorbance and the transmittance, in Van der Waals heterostructures of graphene monolayer on 2D transition metal dichalcogenide (GrTMD) substrate here. We obtain the gapped bands with a Rashba spin-orbit coupling dependent pseudo Zeeman field due to the interplay of substrate induced interactions. This enables us to obtain an expression of the dielectric function in the finite doping case ignoring the spin-flip scattering events completely. We observe that the relative strength of screening is nearly a constant with relative to the changes in the the carrier density. The stronger confinement capability of GrTMD Plasmon(tunable) compared to that of standalone graphene is a major outcome of our analysis; the plasmon dispersion yields the $q^{2/3}$ behavior and not the well known $q^{1/2}$ behavior. We also find that the absorbance and the transmittance are increasing functions of the frequency and the gate voltage. This outcome is useful for devices utilizing photoconductivity and the photo-electric effect.

**Main Text** There has been an intense focus on 2D materials[1-9] ever since the van der Waals heterostructures[10-14], assembled from atomically thin layers of graphene, hexagonal boron nitride and other related materials, have created a new paradigm in materials science revealing unusual properties and new phenomena. These structures not only exhibit an incredible (nano) technological possibility, they also provide a fascinating dais for theoretical explorations through the handling of their incarcerated electronic systems. For example, the higher degree of incarceration and longer lifetimes of graphene plasmons, accessed in graphene encapsulated boron nitride crystals [15-17], have stimulated intense efforts to study such collective excitations triggered by the prospect of paving the way for architecting nano-photonic and nano-electronic devices and components. Bridging the whole spectral range from the mid-infrared to the terahertz (THz) band [18], the phonon, the ultra-confined, long-lived ($\geq 500$ fs) plasmons (wavelength nearly 100 times less than the free-space excitation wavelength) [1,19], and the plasmon-phonon polariton modes are accommodated in such structures and corresponding properties are inter-weaved. The gate voltage and/or added impurities facilitate efficacious manipulation of these effects [20,21].

In this communication we study the interesting and useful possibilities related to the dielectric properties of graphene (Gr) monolayer on 2D transition metal dichalcogenide (TMD) substrate. The possibilities follow on the heels of the engineering of the enhanced spin-orbit coupling (SOC) in graphene through interfacial effects via coupling to the substrate. The substrate-driven interaction (SDI) terms are, in fact, (i) the orbital gap related to the transfer of the electronic charge from Gr to TMD, (ii) the sub-lattice-resolved, giant intrinsic SOCs due to the hybridization of the carbon orbitals with the d-orbitals of the transition metal, and (iii) the extrinsic Rashba spin-orbit coupling (RSOC) that allows for external tuning of the band gap in graphene and connects the nearest neighbors with spin-flip. These interactions are time-reversal invariant and absent by inversion symmetry in isolated pristine, pure graphene monolayer. The graphene layer is also exchange($M$) coupled with the magnetic impurities, e.g. Fe atoms deposited to the graphene surface. A direct, functional electric field control of magnetism at the nano-scale is needed for the effective demonstration of our results related to the exchange-field dependence. The magnetic multi-ferroics, like $BiFeO_3$ (BFO) have piqued the interest of the researchers world-wide with the promise of the coupling between the magnetic and electric order parameters. The SOC interactions and the exchange field are in the band and do not act as scatterers here. We have not considered the intrinsic RSOC for the following reason: Unlike conventional semiconducting 2D electron gases, in which the Rashba coupling is modeled as $\alpha(\delta k_y \sigma_x - \delta k_x \sigma_y)$ where $\sigma's$ are the Pauli matrices, the Rashba coupling in graphene does not depend on the momentum components ($\delta k_x, \delta k_y$). The reason is that Rashba coupling is proportional to velocity, which is constant for mass-less Dirac electrons in graphene.

We obtain the gapped bands [22] with a RSOC-dependent pseudo Zeeman field due to the interplay of SDIs. It may be mentioned that the pseudo-spin alignment is also possible in graphene through a pseudo Zeeman field induced by mechanical deformation related vector potential. The field couples with different signs to states in the two valleys. Unlike this, our effective Zeeman field couples with same sign but different manner to the states in the valleys. This field encourages the spin precession due to the effective magnetic field in the system over the spin-flip scattering of electrons due to momentum scattering. Thus, the field enables us to obtain an expression of the dielectric function in the finite doping case ignoring the spin-flip scattering events completely. We find that the graphene on TMDs [22] is gapped at all possible exchange field values. On account of the strong spin-orbit coupling, the system acts as a quantum spin Hall insulator for $M = 0$. As the exchange field ($M$) increases, the band gap narrowing takes place followed by its recovery. The essential features of the bands, apart from the particle-hole symmetry, are (i) opening of an orbital gap due to the effective staggered potential, (ii) spin splitting of the bands due to the Rashba spin-orbit coupling and the exchange coupling, and (iii) the band gap narrowing and widening due to the many-body effect and the Moss-Burstein effect [23] respectively. The latter is due to the enhanced exchange effect. The exchange field $M$ arises due to proximity coupling to ferromagnetic impurities, such as depositing Fe atoms to the graphene surface. Our plot[22] for the Dirac point K shows that as the exchange field increases the relevant band gap between the spin-down conduction band and the spin-up valence band gets narrower followed by the gap recovery and the gap widening. As regards $MoY_2$, we find that there is Moss-Burstein (MB) shift only and no band narrowing. Therefore, the exchange field could be used for the efficient tuning of the band gap in graphene on TMD. The shift due to the MB effect is usually observed due to the occupation of the higher energy levels in the conduction band from where the electron transition occurs instead of the conduction band minimum. On account of the MB effect, optical band gap is virtually shifted to high energies because of the high carrier density related band filling. This may occur with the elastic strain as well. Thus, studies are required to establish the simultaneous effect of the strain field and the carrier density on optical properties of Gr-TMD. We note that the band gap narrowing and the Fermi velocity $v_F$ renormalization, both, in Dirac systems, are essentially many body effects. Our observation of the gap narrowing in graphene on $WSe_2$, thus, supports the hypothesis of $v_F$ renormalization[24]. Furthermore, (i) the direct information on the gap narrowing and the $v_F$ renormalization in graphene can be obtained from photoemission, which is a potent probe of many body effects in solids, and,(ii) as already mentioned, new mechanisms for achieving direct electric field control of ferromagnetism are highly desirable in the development of functional magnetic interfaces.

The plasmons are defined as longitudinal in-phase oscillation of all the carriers driven by the self-consistent electric field generated by the local variation in charge density. This collective density oscillations of a doped graphene sheet (Dirac Plasmons) are distinctively different($n^{1/4}$dependence) from that ($n^{1/2}$dependence) of the conventional plasmons with respect to the carrier density (n) dependence. The former as well as the latter ones exhibit $q^{1/2}$ dependence as is well-known[25,26,27]. The broad reviews on graphene plasmonics with particular emphasis on the excitations in epitaxial graphene and on the influence of the underlying substrate in the screening processes could be found in refs.[18,28]. The great interest[1-30] evinced by the material science community in recent years in the graphene plasmons is linked to the facts that (i) the propagation of this mode has been directly imaged in real space by utilizing scattering-type scanning near-field optical microscopy[20,21], (ii) the graphene plasmon is highly tunable and shows strong energy confinement capability[29], and (iii) the graphene plasmons strongly couple to molecular vibrations of the adsorbates, polar phonons of the substrate, and so on, as they are very sensitive to the immediate environment [30]. Our results for Gr-TMD plasmon are rather remarkable in this context: At first, we note that using Maxwell's equations with appropriate boundary conditions, the plasmon dispersion can be obtained by solving the equation $\varepsilon + i(q/2\omega)\sigma(q,\omega) = 0$ where the permittivity $\varepsilon = (\varepsilon_{substrate} + \varepsilon_0)/2$, $\sigma(q,\omega)$ is the optical conductivity, $q$ is a wave vector, and $\omega$ is the angular frequency of the incident monochromatic optical field. Because of the finite scattering rate, $q = q_1 + iq_2$ has to be a complex variable with $q_2 \neq \mathbf{0}$ for the above equation to be valid. In view of the fact that the speed of light $c$ in vacuum is 300 time higher than the Fermi velocity of graphene, we have ignored the retardation effect in the equation above. We, however, have obtained the plasmon dispersion within the RPA by finding the zeros of the dielectric function, for a finite chemical potential, including the full dispersion of graphene on TMD[22]. This is essentially same as using the equation $\varepsilon + i(q/2\omega)\sigma(q,\omega) = 0$. We replace $\sigma(q,\omega)$ by $\sigma_{RPA}(q,\omega) = (ie^2q^{-2}\omega)\chi(q,\omega)$ where $\chi(q,\omega) = \chi_1(q,\omega) + i\ \chi_2(q,\omega)$ is the dynamical polarization function. It may be mentioned that the most general expression of the dynamical polarization at finite temperature, chemical potential, impurity rate, quasi-particle gap, and magnetic field was presented by Pyatkovskiy and Gusynin [31] several years ago. The dispersion, we obtain, is essentially that of a gapped electron-hole system with the gap being dependent on the valley and spin indices as well as on the exchange interaction($M$). The gap magnitude is found to be decreasing function of $M$ over a broad range of values (0−1meV). The band energies get spin-split beyond $M = 0.9$ meV. This is perhaps an indication of the system cross-over/evolution to Elliot-Yafet [32] spin relaxation, via the route of the greater occurances of the spin-flip scattering events, from a D'yakonov-Perel' [33] type of spin-relaxation complaint environment ($M < 0.8$ meV). It may be noted that whereas in the former the inversion symmetry is retained, in the latter this symmetry is broken. Now with, say, hole doping, the Fermi surface shifts to a lower energy. As a result the inter-band transitions with transition energy below twice the Fermi energy become forbidden, and it leads to a decrease in higher frequency inter-band absorption. At the same time, the lower frequency free carrier absorption (i.e. intra-band transition) increases dramatically. Therefore, for simplicity, we considered the intra-band transitions only, ignoring the spin-flip mechanism completely. Since we focus on the long wavelength plasmons here, we neglect transitions between two Dirac nodes located at different momenta. We find that there is only one collective mode and it corresponds to charge plasmons. We find that plasmon wavelength and graphene lattice constant(*a*) ratio ($\lambda$) as a function of frequency (*f*) is given by $\lambda = K(n) f^{-3/2}$ where $K(n) \sim C\, n^{3/4}$. For $n \sim 10^{16}$ m$^{-3}$, and $a = 2.8\ 10^{-10}$m, we find $a\, K(n) \sim 10^{12}$ m-Hz$^{3/2}$. This leads to the plasmon wavelength as 1μ-m at THz and $10^{-3}$μ-m at the mid infrared spectral range. Thus, in our case the plasmonic dispersion relation is of the form $f \sim \mathrm{const.}\ n^{1/2} q^{2/3}$. In comparison, for the standalone graphene sheet, the plasmon wavelength is $J(n) f^{-2}$ where $J(n) \sim C_1 n^{1/2}$. For the same value of the carrier density we find $J(n) \sim 10^{22}$ m-Hz$^2$. This leads to the plasmon wavelength as 10 mm at THz and 1μ-m at the mid infrared spectral range. The stronger confinement capability of Gr-TMD plasmon is obvious from above. It is gratifying to see that a finite chemical potential applied to a graphene sheet provides a conduction band for the electrons, allowing for plasmons supported by the graphene on TMD. Unlike in standard semiconductors where the carrier type is fixed by chemical doping during the growth process, the Fermi level in graphene can be continuously driven (tuned) between the valence and conduction bands simply by applying a gate voltage, i.e. by electrostatic doping [34]. The tunability aspect of Plasmon frequency is displayed in figure 1(a). We have shown the 2D plots of the plasmon frequency in arb.unit as a function of the gate voltage at different values of the exchange field. We have taken care not to assign larger values ($M > 0.1$meV) of the exchange field which may trigger spin-splitting ignored in our analysis. The plasmon frequency increases with the increase in the absolute value of the gate voltage at a given value of the exchange field. However, the frequency shows a slight red-shift followed by a much larger blue-shift at a given gate voltage with $M$. The reason being the application of an electric field alters the carrier density so does the magnetic exchange interaction.

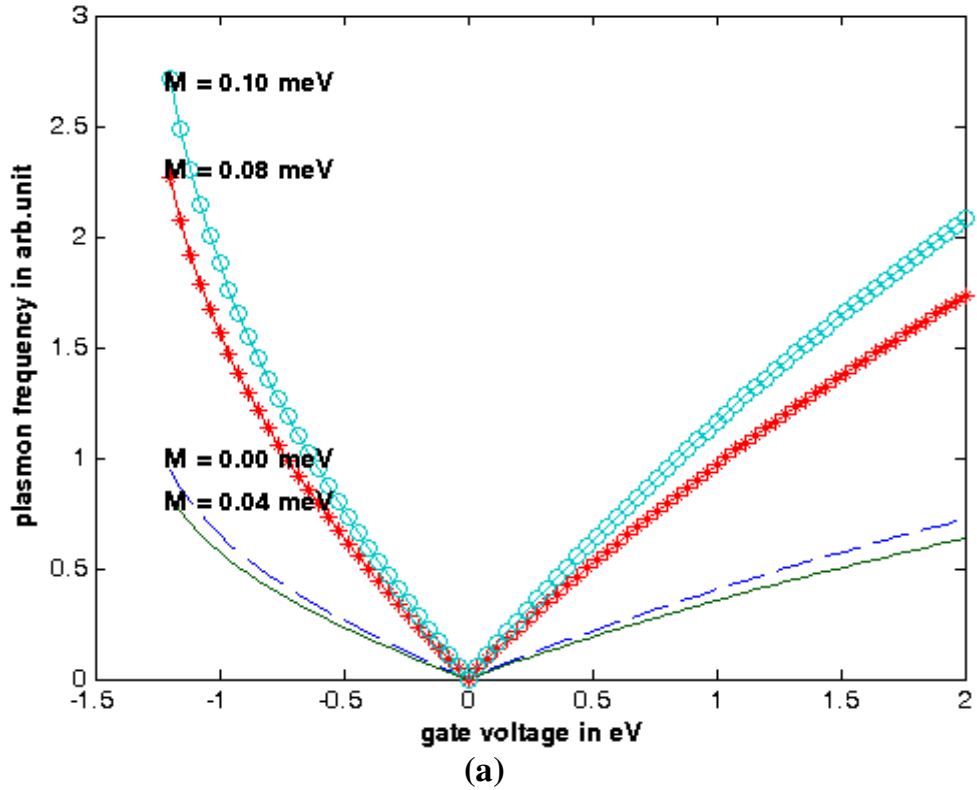

(a)

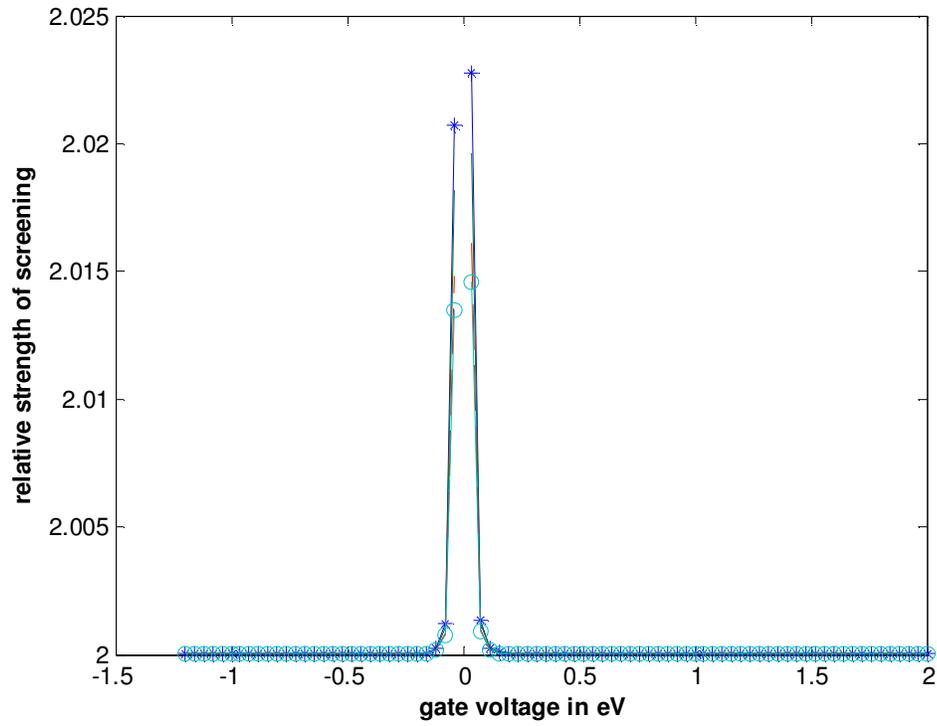

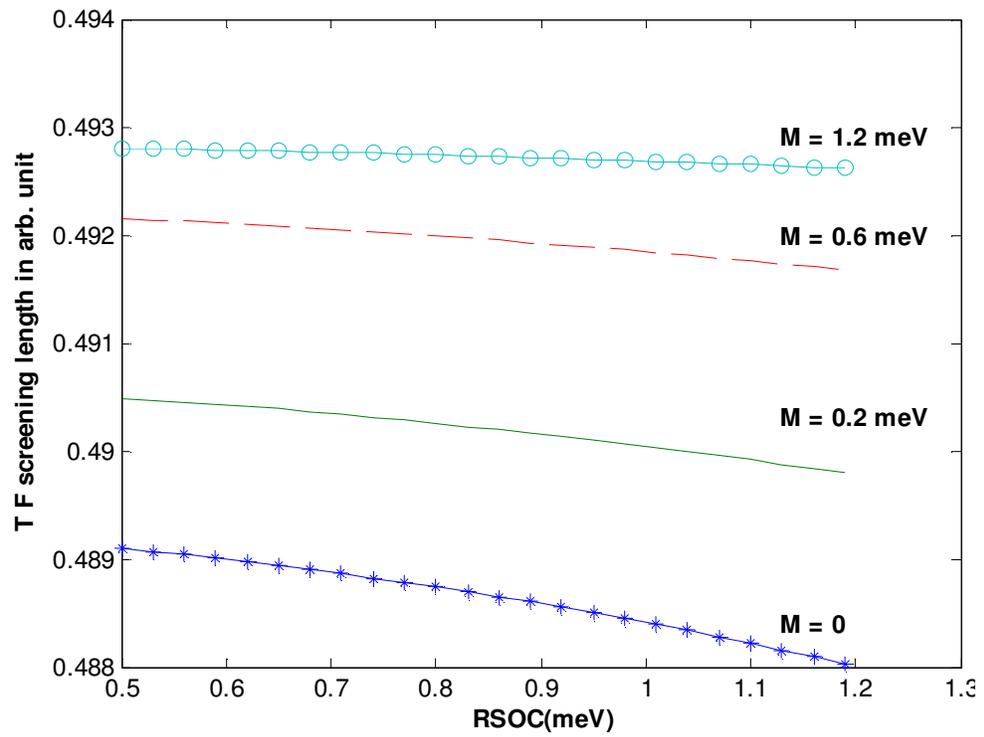

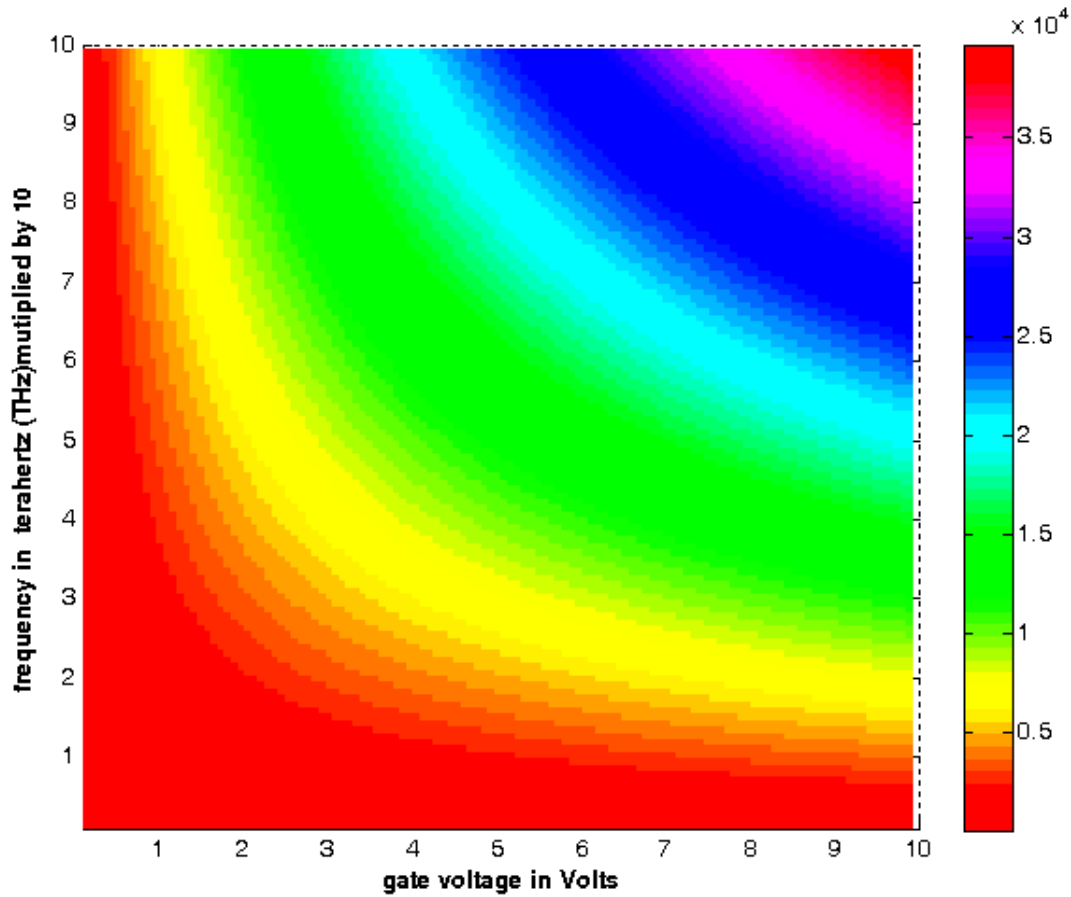

**Figure1.** (a) The 2D plots of the plasmon frequency in arb.unit as a function of the gate voltage. The lowermost corresponds to the exchange field $M = 0.04$ meV, the two curves in the middle to M=0 and $M = 0.08$ meV, and uppermost to $M = 0.10$ meV. (b) The 2D plots of the relative strength of screening as a function of the gate voltage in eV. (c) The plots of the Thomas-Fermi (TF) screening length in arbitrary unit as a function of the RSOC parameter at a given value of the exchange field (M).The lowermost curve corresponds to $M = 0$, and the two curves in the middle to M= 0.2meV and 0.6 meV. The uppermost curve corresponds to M= 1.2 meV. The plots refer to the graphene on $WSe_2$.The chemical potential is assumed to be constant and equal to 6.6743 meV.(d) A contour plot of the real part of the optical conductivity $G_0(\omega, V_g)$ as a function of the frequency and the gate voltage. We find that $G_0(\omega, V_g)$ is increasing function of both the variables.

In conventional 2D electron gases (2DEGs), the Thomas-Fermi wave vector $\kappa$ is generally independent of the carrier density. However, for the pure graphene the screening wave vector is proportional to the square root of the density. Thus, the relative strength of screening ($\kappa / k_F$), where $k_F \sim \sqrt{(\pi n}$ is the Fermi wave-vector, in pure graphene is constant. In the large momentum transfer regime, of course, the static screening increases linearly with wave vector due to the inter-band transition [35]. For Gr-TMD, we obtain the similar result. As shown in Figure 1(b), the relative strength of screening is nearly a constant with relative to the changes in the gate voltage (or, the carrier density) and the exchange field strength; at lower densities the behavior is slightly contrary to what one would expect. We also find that the stronger Rashba coupling (RSOC) has slight foiling effect on the Thomas-Fermi (TF) screening length ($\kappa^{-1}$). This could be seen from Figure 1(c). The RSOC parameter can be tuned by a transverse electric field and vertical strain. Interestingly, from the plots we also notice that the sreening length is greater when the exchange field strength is greater. The plots refer to the graphene on $WSe_2$.The chemical potential is assumed to be constant and equal to 6.6743 meV. The foiling effect in Figure 1(c), over a reasonably broad range of the exchange field values (0–1meV), is possibly an indication of the domination of the spin meandering due to the effective magnetic field corresponding to the pseudo Zeeman term discussed above(Dyakonov-Perel (DP) mechanism [33]) in the system over the spin-flip scattering of electrons due to momentum scattering and spin-orbit splitting of the valence band (Elliot-Yafet (EY) spin relaxation mechanism [32]).The almost nil foiling effect for larger M (M >1meV) indicates establishment of the domination of the Elliot-Yafet mechanism due to alteration in the spin relaxation rate via the introduction of more spin-flip scattering events. We are thus able to show that the characteristics linked to the screening in gated Gr-TMD is nearly insensitive to the substrate induced perturbations and the magnetic impurities.

As regards the absorbance ($A$) and the transmittance ($T$), the latter is given by the equation $T(\omega,V_g) = (1+ G_0(\omega,V_g)/(2c\varepsilon_0))^{-2}$ where $G_0$ is the real part of the optical conductivity. In order to calculate $G_0$ we make use of the relation [36,37]

$$G_0(\omega)=\sigma_0(4\pi/m_e^2\omega^2)\sum_{\sigma,\sigma'} \int (d^2k/(2\pi)^2)(n_\sigma(k)- n_{\sigma'}(k)) \delta\{\hbar\omega + E_\sigma(k) - E_{\sigma'}(k)\}F^{\alpha\beta}_{\sigma,\sigma'}(k),$$

$$F^{\alpha\beta}_{\sigma,\sigma'}(k) = \prod^{\alpha}_{\sigma,\sigma'}(k) \prod^{\beta}_{\sigma,\sigma'}(k), \quad E_\sigma(k), E_{\sigma'}(k) < E_F \qquad (1)$$

Here, the indices $\sigma$ and $\sigma'$ denote the spin and all band quantum numbers for the occupied and empty states respectively, k is the continuous quantum number related to the translational symmetry and restricted to the Brillouin zone, $E_F$ is the Fermi energy, $\sigma_0=(e^2/4\hbar)$ is the frequency-independent universal sheet conductivity of the mass-less Dirac fermions, $m_e$ denotes the electron mass, $\omega$ is the angular frequency of the electromagnetic radiation causing the transition, $n_\sigma(k)$ is the Fermi-Dirac distribution function evaluated at energy $E_\sigma(k)$, and $\prod^{\alpha}_{\sigma,\sigma'}(k) = \langle\sigma',k|p_\alpha^{op}|\sigma,k\rangle$ is the transition matrix element of the α-component of the momentum operator $p_\alpha^{op}$ for a transition from the initial state $|\sigma, k\rangle$ with energy $E_\sigma(k)$ into the final state $|\sigma', k\rangle$ with energy $E_{\sigma'}(k)$. These matrix elements have been obtained from our band structure calculation[22]. We find $G_0^{intraband}(\omega,V_g)$ to be an increasing function of $(\omega,V_g)$ and (approximately) linearly varying with relative to $\omega$ in the limited photon energy range. The consistency of this with the Maxwell's law $\nabla \times B = i\omega\mu\varepsilon(1- (i\sigma/\omega\mu_0\varepsilon_0 n^2))E$, where $B$ and $E$ are the magnetic and electric fields, respectively, $\mu(\mu_0)$ and $\varepsilon(\varepsilon_0)$ are permeability and permittivity, respectively, of the Dirac fermions (free space), $n = \sqrt{(\mu_r\varepsilon_r)}$ the optical index, and $\varepsilon_r$ is the complex relative permittivity, demands that the optical index ($n$) of the Gr-TMD, and, consequently, relative permittivity should be independent of frequency. In view of the pure graphene being known to possess dispersion-less optical index [38,39], the observation that calculated $G_0(\omega,V_g)$ is consistent with the Maxwell's law is not entirely unfounded. A constant optical index, obviously, does not support the optical conductivity value $\sigma_0=(e^2/4\hbar)$. We notice from above that, for the transmission through graphene, only the real part of the optical conductivity is relevant. The reason being the thickness of graphene is several orders of magnitude smaller than an optical wavelength. In Figure 1(d) we have shown a contour plot of the $G_0(\omega,V_g)$ as a function of the frequency and the gate voltage. We find that $G_0(\omega,V_g)$ is increasing function of both the variables. The imaginary part of the optical conductivity $I_0(\omega,V_g)$ is given by the first Kramers-Kronig (KK) relation[40], $I_0(\omega,V_g) = (-2/\pi) P \int_0^\infty \omega G_0(\varepsilon,V_g)(\varepsilon)d\varepsilon/[\varepsilon^2 - \omega^2]$ where $P$ denotes the Cauchy principal value. The second relation is $G_0(\omega,V_g) = (2/\pi) P \int_0^\infty \varepsilon I_0(\varepsilon,V_g)(\varepsilon)d\varepsilon/[\varepsilon^2 - \omega^2]$. We notice that alternatively, if transmission spectra is given, it is possible to obtain the real and the imaginary parts of the ac conductivity.

In conclusion, the Gr-TMD plasmons have unusual properties and offer promising prospects for plasmonic applications covering a wide frequency range, ranging from terahertz up to the visible. Also, we have demonstrated here that the exchange field can be used for efficient tuning of the band gap and the dielectric properties, such as the plasma frequency. The optical conductivity and absorbance, too, can be controlled by tuning of the exchange field.

E-mail for corresponding author: physicsgoswami@gmail.com